\input harvmac
\noblackbox
\newcount\figno
\figno=0
\def\fig#1#2#3{
\par\begingroup\parindent=0pt\leftskip=1cm\rightskip=1cm\parindent=0pt
\baselineskip=11pt
\global\advance\figno by 1
\midinsert
\epsfxsize=#3
\centerline{\epsfbox{#2}}
\vskip 12pt
\centerline{{\bf Figure \the\figno:} #1}\par
\endinsert\endgroup\par}
\def\figlabel#1{\xdef#1{\the\figno}}

\def\np#1#2#3{Nucl. Phys. {\bf B#1} (#2) #3}
\def\pl#1#2#3{Phys. Lett. {\bf B#1} (#2) #3}

\def\prd#1#2#3{Phys. Rev. {\bf D#1} (#2) #3}


\font\cmss=cmss10
\font\cmsss=cmss10 at 7pt
\def\rlx{\relax\leavevmode}
\def\inbar{\vrule height1.5ex width.4pt depth0pt}
\def\IC{\relax\,\hbox{$\inbar\kern-.3em{\rm C}$}}
\def\IN{\relax{\rm I\kern-.18em N}}
\def\IP{\relax{\rm I\kern-.18em P}}
\def\ZZ{\rlx\leavevmode\ifmmode\mathchoice{\hbox{\cmss Z\kern-.4em Z}}
 {\hbox{\cmss Z\kern-.4em Z}}{\lower.9pt\hbox{\cmsss Z\kern-.36em Z}}
 {\lower1.2pt\hbox{\cmsss Z\kern-.36em Z}}\else{\cmss Z\kern-.4em
 Z}\fi}
\def\IZ{\relax\ifmmode\mathchoice
{\hbox{\cmss Z\kern-.4em Z}}{\hbox{\cmss Z\kern-.4em Z}}
{\lower.9pt\hbox{\cmsss Z\kern-.4em Z}}
{\lower1.2pt\hbox{\cmsss Z\kern-.4em Z}}\else{\cmss Z\kern-.4em
Z}\fi}

\def\narrowplus{\kern -.04truein + \kern -.03truein}
\def\narrowminus{- \kern -.04truein}
\def\narrowminussub{\kern -.02truein - \kern -.01truein}

\def\kh{K\"{a}hler}

\def\zbar{\overline{z}}

\def\M{\cal M}
\def\B{\cal B}
\def\b{{\beta}}
\def\a{{\alpha}}

\def\e{{\epsilon}}

\def\r{{\rightarrow}}

\def\frac#1#2{{#1\over #2}}

\def\Bone{{\bf 1}}

\def\IZ{\relax\ifmmode\mathchoice
{\hbox{\cmss Z\kern-.4em Z}}{\hbox{\cmss Z\kern-.4em Z}}
{\lower.9pt\hbox{\cmsss Z\kern-.4em Z}}
{\lower1.2pt\hbox{\cmsss Z\kern-.4em Z}}\else{\cmss Z\kern-.4em
Z}\fi}
\def\IB{\relax{\rm I\kern-.18em B}}
\def\IC{{\relax\hbox{$\inbar\kern-.3em{\rm C}$}}}
\def\ID{\relax{\rm I\kern-.18em D}}
\def\IE{\relax{\rm I\kern-.18em E}}
\def\IF{\relax{\rm I\kern-.18em F}}
\def\IG{\relax\hbox{$\inbar\kern-.3em{\rm G}$}}
\def\IGa{\relax\hbox{${\rm I}\kern-.18em\Gamma$}}
\def\IH{\relax{\rm I\kern-.18em H}}
\def\II{\relax{\rm I\kern-.18em I}}
\def\IK{\relax{\rm I\kern-.18em K}}
\def\IP{\relax{\rm I\kern-.18em P}}

\font\cmss=cmss10 \font\cmsss=cmss10 at 7pt
\def\IR{\relax{\rm I\kern-.18em R}}
\def\pbar{\bar{\p}}

\def\1{{\bf 1}}
\def\3{{\bf 3}}
\def\7{{\bf 7}}
\def\6{{\bf 6}}
\def\2{{\bf 2}}
\def\8{{\bf 8}}

\def\bbar{{\bar b}}

\def\dbar{{\bar d}}

\def\pbar{{\bar p}}
\def\qbar{{\bar q}}
\def\rbar{{\bar r}}
\def\sbar{{\bar s}}

\def\tM{\widehat {\cal M}}
%

%
%
\def\eqnn#1{\xdef #1{(\secsym\the\meqno)}\writedef{#1\leftbracket#1}%
\global\advance\meqno by1\wrlabeL#1}
\def\eqna#1{\xdef #1##1{\hbox{$(\secsym\the\meqno##1)$}}
\writedef{#1\numbersign1\leftbracket#1{\numbersign1}}%
\global\advance\meqno by1\wrlabeL{#1$\{\}$}}
\def\eqn#1#2{\xdef #1{(\secsym\the\meqno)}\writedef{#1\leftbracket#1}%
\global\advance\meqno by1$$#2\eqno#1\eqlabeL#1$$}

\lref\rreid{M. Reid, Math. Ann. {\bf 274} (1987) 329.}
\lref\rcandelas{P. Candelas, L. Parkes, P. Green and X. de la Ossa, 
\np{359}{1991}{21}.}
\lref\rps{J. Polchinski and A. Strominger, hep-th/9510227, \pl{388}{1996}{736}.}
\lref\rGH{P. Griffiths and J. Harris, {\it Principles of Algebraic Geometry}, 
(John Wiley and Sons, 1978).}
\lref\rWNSP{E. Witten, hep-th/9604030.}
\lref\rWI{E. Witten, Nucl. Phys. {\bf B202} (1982) 253.}
\lref\rPKT{P. K. Townsend, Phys. Lett. {\bf B350} (1995) 184.}
\lref\rWSD{E. Witten, Nucl. Phys. {\bf B443} (1995) 85.}
\lref\rAS{A. Strominger, Nucl. Phys. {\bf B451} (1995) 96.}
\lref\rBSV{M. Bershadsky, V. Sadov, and C. Vafa, hep-th/9511222.}
\lref\rVFT{C. Vafa, hep-th/9602022, \np{469}{1996}{403}.}
\lref\rVWOL{C. Vafa and E. Witten, Nucl. Phys. {\bf B447} (1995) 261.}
\lref\rDLM{M. Duff, J. Liu, and R. Minasian,  Nucl. Phys. 
{\bf B452} (1995) 261.}
\lref\rBB{K. Becker and M. Becker, 
hep-th/9605053, \np{477}{1996}{155}.}
\lref\rfluxquant{E. Witten, hep-th/9609122, J. Geom. Phys. {\bf 22} (1997) 1.}
\lref\rgreensav{M. B. Green and S. Sethi, hep-th/9808061, \prd{59}{1999}{46006}.}
\lref\rnaka{H. Nakajima, ``Lectures on Hilbert Schemes of Points on Surfaces,''
to appear.}
\lref\rsvw{S. Sethi, C. Vafa and E. Witten, hep-th/9606122, \np{480}{1996}{213}.}
\lref\rmd{K. Dasgupta and S. Mukhi, hep-th/9612188, \pl{398}{1997}{285}.}
\lref\rbs{M. Bershadsky and V. Sadov, hep-th/9703194, \np{510}{1998}{232}.}
\lref\rsenorien{A. Sen, hep-th/9605150, \np{475}{1996}{562}.}
\lref\rsenftheory{A. Sen, hep-th/9709159, \prd{55}{1997}{7345}.}
\lref\rstrom{A. Strominger, Nucl. Phys. {\bf B274} (1986) 253.}
\lref\rrocek{M. Ro\v cek
in {\it Essays on Mirror Manifolds} vol. 1, ed. by S.-T. Yau, (International 
Press, 1992); S. Gates, C. Hull and M. Ro\v cek, \np{248}{1984}{157}.}
\lref\rkirit{E. Kiritsis, C. Kounnas and D. L\"ust, hep-th/9312143, in
{\it Mirror Symmetry II}, ed. by B. Greene and S.-T. Yau, (AMS and International
Press, 1997).}
\lref\rDJM{K. Dasgupta, D. P. Jatkar and S. Mukhi, hep-th/9707224, \np{523}{1998}{465}.}
\lref\rBUSH{T. Buscher, Phys. Lett. {\bf B194} (1987) 59; {\bf B201} (1988) 466.}
\lref\rKKL{E. Kiritsis, C. Kounnas and D. L\"ust, hep-th/9308124,
Int. J. Mod. Phys. {\bf A9} (1994) 1361.}
\lref\rBBO{E. Bergshoeff, H. J. Boonstra and T. Ortin, 
hep-th/9508091, \prd{53}{1996}{7206}.}
\lref\rBHO{E. Bergshoeff, C. Hull and T. Ortin, hep-th/9504081, \np{451}{1995}{547}.}
\lref\rWdynm{E. Witten, hep-th/9503124, Nucl. Phys. {\bf B443} (1995).}
\lref\rgukov{S. Gukov, C. Vafa and E. Witten, hep-th/9906070.}
\lref\rgreene{B. Greene and R. Plesser, \np{338}{1990}{15}.}
\lref\rblumen{R. Blumenhagen, R. Schimmrigk and A. Wisskirchen, 
hep-th/9609167, \np{486}{1997}{598}.}
\lref\rdewit{B. de Wit, D. Smit and N. Hari Dass, \np{283}{1987}{165}.}
\lref\rhull{C. Hull in {\it Superunification and Extra Dimensions}, ed. by
R. D'Auria and P. Fr\'e, (World Scientific Publishing, 1986).}
\lref\rshioda{T. Shioda and H. Inose in {\it Complex Analysis and Algebraic
Geometry}, ed. by W. Baily, Jr. and T. Shioda, (Cambridge University Press,
1977).}
\lref\rbtwo{M. Bershadsky, M. Johansen, T. Pantev and V. Sadov, hep-th/9701165,
\np{505}{1997}{165}.}
\lref\rgopak{R. Gopakumar and S. Mukhi, hep-th/9607057, \np{479}{1996}{260}.}
\lref\rzurab{Z. Kakushadze, G. Shiu and H. Tye, hep-th/9804092, \np{533}{1998}{25};
Z. Kakushadze, hep-th/9904211.}
\lref\rplesser{D. R. Morrison and R. Plesser, hep-th/9810201, A.T.M.P. {\bf 3} (1999) 1.}
\lref\rvafw{C. Vafa and E. Witten, hep-th/9505053, \np{447}{1995}{261}.}
\lref\rmin{M. Duff, J. Liu and R. Minasian, hep-th/9506126, \np{452}{1995}{261}. }

\lref\rVDT{C. Vafa, Nucl. Phys. {\bf B273} (1986) 592.}
\lref\rVWDT{C. Vafa and E. Witten, J. Geom. Phys. {\bf 15} (1995) 189.}
\lref\rwitcos{E. Witten, Mod. Phys. Lett. {\bf A10} (1995) 2153.}
\lref\rPKT{P. K. Townsend, Phys. Lett. {\bf B350} (1995) 184.}
\lref\rWSD{E. Witten, Nucl. Phys. {\bf B443} (1995) 85.}
\lref\rJAT{D. P. Jatkar and S. K. Rama, 
hep-th/9606009.}
\lref\rTS{A. A. Tseytlin,  hep-th/9602064. }
\lref\rugw{E. Witten, 
Nucl. Phys. {\bf B463} (1996) 383.}
\lref\rSE{A. Sen,  hep-th/9604070.}
\lref\rMVFT{D. Morrison and C. Vafa,
hep-th/9602114, hep-th/9603161.}
\lref\rbottu{R. Bott and L. W. Tu, {\it Differential Forms In Algebraic
Topology} (Springer-Verlag, 1982).}
\lref\rBSV{M. Bershadsky, V. Sadov and C. Vafa, Nucl. Phys. 
{\bf B463} (1996) 398.}
\lref\rL{M. Li, Nucl. Phys. {\bf B460} (1996) 351.}
\lref\rMD{M. Douglas, ``Branes within Branes'', hep-th/9512077.}
\lref\rVI{C. Vafa, Nucl. Phys. {\bf B463} (1996) 435.}
\lref\rMH{M. Green, J. Harvey and G. Moore, ``I-Brane Inflow and 
Anomalous Couplings on D-branes'', hep-th/9605033.}
\lref\rEZ{E. Zaslow, Comm. Math. Phys. {\bf 156} (1993) 301.}
\lref\rLD{M. Douglas and M. Li, ``D-brane Realization of Super Yang-Mills
Theory in Four-Dimensions'', hep-th/9604041. }
\lref\rSW{N. Seiberg and E. Witten, ``Comments on String Dynamics in
Six-Dimensions'', hep-th/9603003.}
\lref\rCF{P. Candelas and A. Font, ``Duality Between the Webs of
Heterotic and Type II Vacua'', hep-th/9603170.}
\lref\rWP{E. Witten, ``Phase Transitions In M-Theory And F-Theory,''
hep-th/9603150.}

\Title{\vbox{\hbox{hep-th/9908088}
\hbox{IASSNS--HEP--99/75, NSF-ITP-99-095 }}}
{\vbox{\centerline{M Theory, Orientifolds and $G$-Flux}}}
\centerline{Keshav Dasgupta\footnote{$^1$}{keshav@sns.ias.edu}, 
Govindan Rajesh\footnote{$^2$}{rajesh@sns.ias.edu} and 
Savdeep Sethi\footnote{$^3$}{sethi@sns.ias.edu}}
\vskip 0.1 in
\medskip\centerline{\it School of Natural Sciences}
\centerline{\it Institute for Advanced Study}\centerline{\it
Princeton, NJ
08540, USA}

\vskip 0.5in

We study the properties of M and F theory compactifications to three and four
dimensions with background fluxes. We provide a simple construction of supersymmetric
vacua, including some with orientifold descriptions. These vacua, 
which have warp factors, typically have fewer moduli
than conventional Calabi-Yau compactifications. The mechanism for anomaly cancellation
in the orientifold models involves background RR and NS fluxes. We consider 
in detail an orientifold of $K3\times T^2$ with background fluxes. After a 
combination of T and S-dualities, this type IIB orientifold is mapped to a 
compactification of the $SO(32)$ heterotic string on a non-\kh\ space with torsion.   

\vskip 0.1in
\Date{8/99}

\newsec{Introduction}

The realization that certain string compactifications can be described
in multiple ways has provided a significant improvement in our
understanding of non-perturbative string dynamics. Among the more
interesting compactifications are those with N=1 and N=2
supersymmetry in four dimensions. An intriguing class of such four
dimensional vacua are described by compactifying F theory on a
Calabi-Yau four-fold $\M$ \rVFT. These vacua are naturally related to
compactifications of type IIA string theory and M theory to two
and three dimensions, respectively. The strong coupling limit of type
IIA compactified on $\M$ is well described by the
three dimensional theory obtained by compactifying M theory on the
same four-fold. If the four-fold admits an elliptic fibration with base $\B$, then
we can consider a particular degeneration of this M theory
compactification in which the area of the elliptic fiber shrinks to
zero. In this limit, M theory on the four-fold goes over to type IIB
compactified on the base $\B$ with a varying coupling constant. The coupling
constant is $\tau$ of the elliptic fiber. 
This four-dimensional type IIB vacuum
is known as an F theory compactification.

Among the novel features of these vacua is the need to cancel a
tadpole anomaly. For this class of compactifications, the anomaly is
given by $\chi/24$, where $\chi$ is the Euler characteristic of the
four-fold. If $\chi/24$ is integral, then the anomaly can be
cancelled by placing a sufficient number of spacetime filling branes 
on points of the compactification manifold \rsvw. For type IIA,
the required branes are strings, while in M and F theory, membranes
and $D3$-branes are required, respectively. 

However, there is at least
one other way of cancelling the anomaly in type IIA or M theory, which is
by introducing a background flux for the four-form field strength $G$ \rBB. 
The $G$-flux 
contributes to the membrane tadpole in M theory through the 
Chern-Simons interaction,
\eqn\cs{ \int{ C\wedge G\wedge G.}}
For cases 
where $\chi/24$ 
is not integral,  $G$-flux is actually required to obtain a consistent vacuum. 
In general, the anomaly can be cancelled by a combination of background
flux and branes.  With $n$ background branes,  the tadpole cancellation condition
\eqn\condition{ {\chi\over 24} = {1\over 8 \pi^2} \int{ G\wedge G} + n,}
must be satisfied for type IIA or M theory \rmd. 

Compactifications
with $G$-flux have a number of interesting features, but
have received little attention. The goal of this paper is to explore
some of the properties of these vacua. In the following section, we 
review the results of \rBB\ where conditions  on supersymmetric  
vacua with $G$-flux were derived from eleven-dimensional supergravity.
These conditions are quite difficult to satisfy. As a consequence, the presence of 
$G$-flux typically freezes some of the geometric moduli of the four-fold in a way
that we describe. 

Some of these M theory vacua can be lifted to 
four-dimensional N=1 F theory vacua. 
The corresponding F theory vacua 
have background fluxes of two kinds. The first kind involves non-zero  
NS and RR three-form field strengths,  
denoted $H$ and $H'$ respectively.
The three-form fluxes contribute to the $D3$-brane tadpole through the 
type IIB supergravity interaction,
\eqn\iib{ \int{ D^+\wedge H\wedge H'},}
where $D^+$ is the four-form gauge field.   In the second kind of background, 
some of the seven-brane gauge fields have non-zero instanton number. In F
theory, this possibility has been discussed in \rbtwo. These
instantons contribute to the $D3$-brane tadpole through
the seven-brane world-volume coupling,
\eqn\nextiib{\int D^+ \wedge F \wedge F,}
where $F$ is the field strength for the seven-brane gauge-field.

F theory is a useful description of these compactifications only when the base $\B$ is
large compared to the string scale. However, for special choices of four-fold $\M$,  
F theory can be related to a type IIB orientifold which is a complete perturbative 
string theory \rsenftheory. In  turn, some of these IIB orientifolds 
can be related to type I 
compactifications via $T$-duality. Largely for their simplicity, orientifolds of
tori are most commonly studied. In conventional models, tadpole cancellation
is achieved by adding branes: either $D9$ and $D5$-branes or $D7$ and $D3$-branes 
depending on the choice of orientifold action. 
In four dimensions, the possibility of using the type IIB interactions 
\iib\ and \nextiib\ to cancel the $D3$-brane tadpole 
suggests that novel orientifolds should exist with backgrounds involving
$H$ and $H'$-fluxes and gauge-field instantons. The possible types of 
orientifold are then classified by the choice of $C$-flux and $G$-flux
in M theory on $\M$. 

In section three, we present examples of vacua with
$G$-flux, including a simple orbifold construction and an example 
where $G$-flux is required. In the final section, we consider an 
example of a type IIB orientifold 
with constant background $H$ and $H'$-fluxes. 
Depending on the choice of background flux, the 
model has either N=1 or N=2 spacetime 
supersymmetry.
The compactification space $\B$ is conformal to $K3\times T^2$. This orientifold
is related to F theory on $K3\times K3$, some aspects of which have 
been discussed in \rbs. 

However, 
our interest is largely with a heterotic dual of this orientifold. By T-dualizing 
along the $T^2$, we map the IIB orientifold to a type
I compactification on a new space $\B'$ with non-zero $H'$-torsion. 
A further S-duality 
turns the type I vacuum
into a perturbative $SO(32)$ heterotic vacuum with non-zero $H$-flux. 
The non-\kh\ space $\B'$ is no longer conformally Calabi-Yau. This is a 
concrete example, possibly the first, of a four-dimensional string compactification
with torsion. 

There are a number of promising directions to explore. 
For example, associated to each of the type I/heterotic
supergravity solutions
is a  world-sheet conformal field theory.
Finding ways of constructing and analyzing these conformal field theories 
is a potentially rewarding enterprise.
There should be analogues in this more general class of 
compactifications of  phenomena
associated with Calabi-Yau compactifications, such as mirror symmetry and
its $(0,2)$ cousin \refs{\rgreene, \rblumen}. There are also
intriguing connections between these solutions and the work of \refs{\rps, \rcandelas,
\rreid}. During the completion of this project, an interesting paper appeared \rgukov\
with some overlap with section 2.

\newsec{Supersymmetry and $G$-flux}
\subsec{$C$-field instantons}

We begin by recalling the results of \rBB\ for M theory
compactified on an eight-dimensional Calabi-Yau manifold $\M$. Let $M_{pl}$ denote
the eleven-dimensional Planck scale. At leading order in a momentum expansion,
the M theory effective action is given by eleven-dimensional supergravity. A product
metric on $ \IR^3 \times \M$ is a
solution to the supergravity equations of motion when the metric for the internal 
space $\M$ is Ricci flat. Let us parametrize $\IR^3$ by coordinates $x^\mu$ where 
$\mu=0,1,2$
and the internal space $\M$ by complex coordinates $y^a$ where $a=1,\ldots,4$. 

At next order in the derivative expansion, there are terms
with eight derivatives which are therefore suppressed by six additional 
powers of $M_{pl}$. Among these terms is an interaction of the form 
\refs{\rvafw, \rmin}, 
\eqn\tadpole{ -\int C \wedge X_8(R),}
where $X_8$ is an eight-form constructed from curvature tensors. This term
induces a tadpole for the $C$-field. A way to cancel the tadpole is to turn on
a non-trivial $G$-flux. The metric is then modified from a simple product and
takes the form:    
\eqn\lowest{ ds^2 = e^{-\phi(y)}\eta_{\mu\nu} dx^\mu dx^\nu + e^{{1\over 2}
\phi(y)} g_{a\bbar} dy^a dy^\bbar.}
The metric $g$ for $\M$ is \kh\ and Ricci flat.
Let us call the warped internal space $\tM$.
The space $\tM$ is therefore conformal to a Calabi-Yau manifold.  

There is also a non-vanishing four-form field strength 
$G$ on $\M$ which satisfies the conditions:
\eqn\beckerstuff{G_{abcd} = G_{abc\dbar}=0, \qquad g^{c\dbar}G_{a\bbar 
c\dbar} =0.}
The only other non-vanishing component of $G$ is given in terms of the warp
factor, 
\eqn\spacialG{ G_{\mu\nu\rho a} = 
\e_{\mu\nu\rho}\, \partial_a e^{-{3\over 2}\phi}.}
Lastly, the warp factor satisfies the equation:
\eqn\warp{ \Delta (e^{{3\phi\over 2}}) = 
\ast \left\{ 4\pi^2 X_8 - {1\over 2} G\wedge G  - 4\pi^2 \sum_{i=1}^n
\, \delta^8 (y-y_i) \right\}.}
The Laplacian and the Hodge star operator in \warp\ are defined with respect to $g$. 
We have included the possibility of $n$ membranes located at the points $y_i$ 
on $\tM$. The 
combination of $G$-flux and membranes must satisfy \condition.

The first condition 
in \beckerstuff\ tells us that $G$ is a $(2,2)$ form
on $\M$. Dirac quantization requires that the cohomology 
class $[G/\pi]$ be
an element of $H^{(2,2)} (\M, \IZ)$. If $\chi/24 \in \IZ$ then $[G/2\pi]$ is an 
integer
cohomology class \rfluxquant. The second condition in \beckerstuff\ is the 
analogue of the
following condition for a two-form field strength $F$:
\eqn\analogue{ g^{a\bbar}F_{a\bbar} = 0.}
In the two-form case, \analogue\ implies that $F$ is anti-self-dual because,
\eqn\twocase{ \eqalign{ (* F)_{a\bbar} &= \e_{ac\bbar \dbar} F^{c\dbar} 
 = (g_{a\bbar} g_{c\dbar} - g_{a \dbar} g_{c \bbar} ) F^{c\dbar} 
 = - F_{a \bbar}.}}
In the four-form case, we can again express the epsilon tensor in the following
way:
$$  \e_{abcd \pbar\qbar\rbar\sbar} = g_{a\pbar} g_{b\qbar} g_{c\rbar} g_{d\sbar} \,
\pm \, {\rm permutations.}$$
In much the same way as the two-form case \twocase, the 
conditions \beckerstuff\ imply that,
\eqn\sd{ G = *G, }
where the Hodge star acts on the internal eight manifold with metric $g$.
This M theory 
background therefore involves an abelian `instanton' of the $C$-field. Lastly, let us 
rephrase the second condition \beckerstuff\ in terms of the \kh\ form of $\M$,
\eqn\kahler{J \equiv i g_{a\bbar} dz^a\wedge dz^\bbar.} 
In terms of $J$, the second condition states that the self-dual $G$-field is
primitive:
\eqn\primitive{ J \wedge G = 0.}
This condition actually means that $G$ is a singlet of the $sl_2$-algebra generated
by $J$, its adjoint $ J^\dagger$ and their commutator $[J, J^\dagger]$. See, for 
example, \rGH. 

\subsec{Compactifications with extended supersymmetry}

For compactifications on $\IR^3 \times \M$, we decompose a $32$ component 
Majorana-Weyl spinor under $SO(2,1)\times SO(8)$ in the following way, 
\eqn\decomp{ {\bf 32} = ({\bf 2}, {\bf 8_s})\oplus ({\bf 2}, {\bf 8_c}). } 
For spaces $\M$ with holonomy $SU(4)$ and not a proper subgroup, we can further
decompose the $SO(8)$ representations under $SU(4)$:
\eqn\defurther{\eqalign{{\bf 8_s} & = \6 \oplus \1 \oplus \1, \cr
{\bf 8_c} & = \bf{4} \oplus \bar{{\bf { 4}}}. }}
The two singlets in ${\bf 8_s}$ give the four real unbroken supersymmetries needed for
a model with N=2 supersymmetry. 

If the holonomy of  $\M$ is a proper subgroup of $SU(4)$,
the theory may have extended supersymmetry. See \rplesser\ for a discussion
of the possible holonomies of an eight manifold. We will discuss two examples which 
appear in later discussion. The first is compactification on a hyper\kh\ manifold 
with holonomy $Sp(2)$. On decomposing the ${\bf 8_s}$ and ${\bf 8_c}$
representations, we find        
\eqn\hyperkh{\eqalign{{\bf 8_s} & = \bf{5} \oplus \1 \oplus \1 \oplus \1, \cr
{\bf 8_c} & = \bf{4} \oplus {\bf {4}}. }}
Therefore, this compactification can have N=3 supersymmetry. Since the space is
hyper\kh, there is a $\IP^1$ of choices of complex structure.  To see this,
note that we can construct a complex structure tensor ${J^i}_j$ from any complex
covariantly constant spinor $\eta$ in the usual way:
\eqn\complx{ {J^i}_j = i \eta^\dagger {\gamma^i}_j \eta. }
The indices $i,j=1,\ldots, 8$ and $\gamma^{i}$ are gamma matrices for $\M$.
For each of the $\IP^1$ of choices for $\eta$, where $\eta$ has norm one, 
there is a corresponding complex structure tensor. 

  The second example 
has holonomy $Sp(1)\times Sp(1)$. In this case, 
\eqn\prodkh{\eqalign{{\bf 8_s} & = (\2,\2) \oplus (\1,\1) 
\oplus (\1,\1) \oplus (\1,\1) \oplus (\1,\1), \cr
{\bf 8_c} & = (\2,\1) \oplus (\2,\1) \oplus (\1, \2) \oplus (\1, \2). }}
This compactification has at most N=4 supersymmetry. There is a $\IP^1\times \IP^1$ of
complex structures. Whether 
there is extended supersymmetry actually depends on the choice of $G$-flux.
To preserve more than N=2 supersymmetry, the $G$-flux must be a primitive $(2,2)$
class with respect to more than a single complex structure. In section three, we
will meet examples of $G$-flux which do not preserve all the supersymmetries of the
compactification manifold.

\subsec{Kaluza-Klein reduction with a warp factor}

In the presence of background fluxes and a warp factor, the counting of
light degrees of freedom is typically quite difficult because the equations obeyed
by the metric and $C$-field fluctuations are coupled. The metric takes the
form, 
\eqn\wmetric{ \pmatrix{e^{-\phi} \eta_{\mu \nu} & 0 \cr 0 & {\widehat g} },}
where ${\widehat g} = e^{{1\over 2}\phi} g$.
Let us begin by considering purely metric deformations. 
Since the metric $g$ is Calabi-Yau, 
infinitesimal deformations
of $g$ are classified in the usual way by elements of $H^{3,1}(\M)$ and 
$H^{1,1}(\M)$.
However, those complex structure deformations that do not keep $G$ a $(2,2)$ class
become massive. Likewise, deformations of the \kh\ structure that do not keep
$G$ primitive become massive. In this way, we generically lose a large number of 
geometric moduli. 

Let us take the generic case where N=2 spacetime supersymmetry is
preserved. The moduli that we wish to count appear in two kinds of multiplet: the first
is the dimensional reduction of a four-dimensional N=1 chiral multiplet. The second is
the reduction of an N=1 vector multiplet. In three dimensions, the vector multiplet can
be dualized to a chiral multiplet containing a dual scalar.
Each surviving deformation in $H^{3,1}(\M)$ gives rise to a chiral multiplet, while each
deformation in
$H^{1,1}(\M)$ gives rise to the scalar field of a vector multiplet. The vector field itself
comes from a $C$-field zero-mode.  

 Let us consider
the effect of just a warp factor on the $C$-field equations.
 Without $G$-flux, 
the $C$-field obeys the free-field equation,
\eqn\freefield{ d \, {\hat *}\, d C = 0. } 
By using the gauge invariance $ C \r C + d \Lambda$, we can demand that $C$ satisfy
\eqn\gauge{ d \, {\hat *}\, C=0.}
Combined with the field equation, this gives the usual condition,
\eqn\free{ \widehat{\Delta} C = 0.}
Decomposing \free\ into spacetime and internal components gives, 
\eqn\decompose{ \left\{\, \partial^\mu \partial_\mu + e^{-\phi} {\widehat 
\Delta_g} 
- {3\over 2} e^{-\phi} \, \,
{\widehat g}^{ab}(\partial_a \phi) \partial_b \, \right\} C =0. }
The last two terms in \decompose\ will look like mass terms from the perspective of the
three-dimensional observer. The second term is conventional and leads to the usual harmonicity
condition on the internal components of $C$. However, the third term is a new 
consequence of the warp factor. 

In the presence of non-trivial background $G$-flux,  equation \freefield\ for a 
fluctuation $\delta C$ becomes:
\eqn\newdecomp{ d \, \hat{*} \, d \, \delta C = - G \wedge d\, \delta C.}
We have set all metric fluctuations to zero in \newdecomp. 
We decompose $\delta C$ into a product of spacetime and internal fields,
\eqn\dC{ \delta C = \psi(x) C^{(3)}(y) + A_\mu(x) C^{(2)}(y),}
where we only consider spacetime multiplets with vector or scalar fields, 
and $C^{(n)}$ is an $n$-form on the eight manifold. We can dispense immediately
with the counting of vector fields $A_\mu(x)$ since each zero-mode of $C^{(2)}$ pairs
with a \kh\ class deformation to give a vector multiplet. 
We only need to count the number of massless modes from $H^{1,1}(\M)$ to count the
number of vector multiplets. 

The final source of moduli are the analogues of Wilson lines for the 
$C$-field. In the absence of $G$-flux, any element of $H^{2,1}(\M)$ gives rise to 
a zero-mode for $C^{(3)}$ and therefore a chiral multiplet.\foot{Note 
that $H^{3,0}(\M)$ is empty for a simply
connected Calabi-Yau.} In the presence of $G$-flux, the conditions on $C^{(3)}$ 
are modified. We can expand the left hand side of  \newdecomp\ as follows,
\eqn\expandnew{ d \, \hat{*} \, d \, \delta C = 
d\, * d \psi \wedge * C^{(3)} + * d \psi \wedge d \, *  C^{(3)}
+ * \psi \wedge d( e^{-3\phi/2} * dC^{(3)}),}
where each Hodge star appearing on the right hand side is with respect to the
{\it unwarped} spacetime and internal metrics.  

Combining \expandnew\ with the right hand side of \newdecomp\ gives the following
set of equations:
\eqn\seq{ \eqalign{ \psi \, G \wedge dC^{(3)} & =0, \cr
 d\psi \wedge G \wedge C^{(3)} & = 0, \cr
 *\, d \psi \wedge d\, * C^{(3)} &=0, \cr
 d\, *  d \psi \wedge * C^{(3)} &= * \psi \wedge d \left( e^{-3\phi/2} \{
* d C^{(3)} -d C^{(3)} \} \right).}}
To satisfy the third equation in \seq, we can fix the gauge by demanding
that,
\eqn\newgauge{d \, *  \delta C = 0.}
This choice differs from the usual gauge fixing condition \gauge\ by an exact
form. As a consequence of \newgauge, $ d\, *  C^{(3)}=0$.
The first equation in \seq\ is a consequence of equation two which
requires that, 
\eqn\newcond{ G \wedge  C^{(3)} =0.}
This condition is the analogue of the primitivity condition \primitive\ for the
metric. 

The right hand side of the final equation in \seq\ must vanish since this term
gives a mass to the spacetime field $\psi$.  Requiring that the perturbation
$dC^{(3)}$ take us to a supersymmetric vacuum implies that 
$ dC^{(3)}$ must be self-dual, which in turn implies that $C^{(3)}$ is harmonic. 
Therefore any element of $H^{2,1}(\M)$ which satisfies
\newcond\ gives rise to a chiral multiplet.

\subsec{Lifting $G$-flux to F theory}

If the eight manifold $\M$ is elliptically-fibered with base $\B$, then by 
shrinking the volume of the fiber, we can lift our M 
theory compactification to a four-dimensional type IIB compactification on
$ \IR^4 \times \B$. Since the power of the warp factor is different for the 
spacetime and internal metric in \wmetric, we might worry that the resulting
four-dimensional metric breaks Lorentz invariance. 

Let us first show that this is not the case. M theory on $T^2$ with area $A$ maps 
to type IIB on a circle $S^1$ with radius proportional to $1/A$. Locally, the
warp factor rescales the metric sending:
$$ A \,\r\, e^{\phi/2} A. $$
However, this corresponds to rescaling the IIB circle metric by
$$ {1\over A^2} \,\r\, e^{-\phi} {1\over A^2}, $$
which is precisely the power needed to obtain a Lorentz invariant four-dimensional
metric.

How does the $G$-flux lift to type IIB? Let us start with the $G_{\mu\nu\rho a}$
component. This component has the form $ d C_{\mu\nu\rho}$, and the this $C$-field
lifts to a component of the four-form $D^+$ of type IIB:
\eqn\firstlift{  C_{\mu\nu\rho} \, \r \, D^+_{\mu\nu\rho\lambda}.}
Note that $\lambda$ is a spacetime index and that $dD^+$ is not required to be 
self-dual. What is required to be self-dual is the combination,
\eqn\Gfive{ F^+ = dD^+  - {1\over 2} B'\wedge H + {1\over 2} B \wedge H'.} 
Therefore the presence of this spacetime $D^+$ field does not imply that there is
a $D^+$ field in the internal space. 

We can divide the remaining $G$-flux
involving $G_{a\bbar c \dbar}$ into two cases. There could be a component of $G$
with no legs along the fiber. This component would map in the following way, 
$$ G_{a\bbar c \dbar} \, \r \, (dD)^+_{a\bbar c \dbar\lambda}. $$
This flux breaks 
four-dimensional Lorentz invariance. By self-duality of $G$, this case also rules
out the possibility of components with two legs along the fiber. 

The remaining possibility is the case where $G$ is locally the product of a 
three-form on $\B$ and a one-form in the fiber. In this case, we can also 
differentiate between two kinds of $G$-flux. We can see this already in the relation
between M theory
on $K3=T^4/\IZ_2$ and the type IIB orientifold of $T^2$ by $\Omega (-1)^{F_L} \IZ_2$,
where $\Omega$  is world-sheet parity \rsenorien. 
M theory on $T^4/\IZ_2$ has $22$ gauge-fields
obtained by reducing the $C$-field on the $22$ forms in $H^2(K3,\IZ)$. The $22$
forms on $T^4/\IZ_2$ are grouped in the following way: there are $16$ twisted
sector $(1,1)$ forms. Each twisted sector form comes from an $A_1$ singularity
so the gauge group is enhanced to $SU(2)$. The $(2,0)$ and $(0,2)$ forms 
descend from $T^4$ as do $4$ more untwisted $(1,1)$ forms. The gauge group is
$SU(2)^{16} \times U(1)^6$.  

In the limit where $A\r0$, the two untwisted sector forms corresponding to
the class of the fiber and its Hodge dual are no longer normalizable. We can
identify the 
remaining $20$ forms with gauge-fields in the orientifold theory in the following
manner: note that the 
action $ \Omega (-1)^{F_L}$ is an element of $SL(2, \IZ)$ given by the matrix,  
\eqn\mat{  \pmatrix{-1 & 0\cr 0 & -1},}
which projects out both $B$ and $B'$. The only components which survive the
projection have a leg along $T^2$.  The $4$ Kaluza-Klein gauge-fields obtained
by reducing $B$ and $B'$ along one-cycles of $T^2$ are then identified with
the $4$ surviving untwisted sector forms of $T^4/\IZ_2$. The $16$ fixed points
coalesce into $4$ groups of coincident $A_1$ singularities. The gauge
group is enhanced from $SU(2)^4 \r SO(8)$. The $SO(8)$ arises in the 
orientifold picture from placing $4$ $D7$-branes at the location of
each $O7$-plane. 

Therefore, if our $G$-flux is localized around a singular fiber of 
the elliptic fibration, 
it will lift to the field strength of a seven-brane gauge-field. The gauge-field
will have non-zero instanton number on the 4-cycle wrapped by the seven-brane. If 
there are multiple seven-branes then the gauge-group can be non-abelian as in the
$T^4/\IZ_2$ example. In this situation, 
the supergravity analysis is incomplete since the
enhanced gauge symmetry is non-perturbative in M theory. From F theory, we certainly
expect the gauge-field to satisfy the non-abelian Donaldson-Uhlenbeck-Yau equation.
Any holomorphic stable vector bundle would then give a supersymmetric solution. 

For the most part, we will consider the last choice for $G$. In this case, we can 
express $G$ locally in terms of a basis 
for one-forms on the torus,
$$ dz = dx + \tau dy \qquad d\zbar = dx + \bar{\tau} dy, $$
in the following way,
\eqn\expandG{ {G\over 2\pi} = dz \wedge \omega - d\zbar \wedge * \omega,}
where $\omega \in H^{1,2}(\B)$.  The flux then lifts to a 
combination of the NS field strength $H$ and RR field
strength $H'$. The field strengths are given in terms of $\omega,$
\eqn\fieldstr{\eqalign{ H &= \omega - * \omega \cr
H' &= \omega \tau - *\omega \bar{\tau}.}}
The anomaly cancellation condition then becomes,
\eqn\newcondition{ {\chi\over 24} = n - \int{ H\wedge H'},}
where $n$ is the number of D3-branes.

The NS field strength $H$ 
and the RR field strength $H'$
are naturally arranged in an $SL(2,\IZ)$ doublet of type IIB supergravity:\foot{
Unfortunately, the natural complex combinations of $H$ and $H'$ are usually 
denoted $G$ and $G^*$ in the literature. To avoid confusion, we have chosen 
the notation $ \Lambda$ and $\Lambda^*$. See \rgreensav\ for an
explanation of the mapping between the $SU(1,1)/U(1)$ and the $SL(2,\IR)/U(1)$ 
parametrizations of the supergravity moduli space.}
\eqn\doublet{ \eqalign{
 \Lambda & = {1\over \sqrt{\tau_2}} \left( H' - \tau H \right) \cr
 \Lambda^* & = {1\over \sqrt{\tau_2}} \left( H' - \bar{\tau} H \right). \cr
}} 
In terms of $\omega$,
$$ \Lambda \sim \sqrt{\tau_2} \, (*\omega) \qquad \Lambda^* \sim  
\sqrt{\tau_2} \, \omega. $$
In a generic F theory compactification, the IIB fields undergo non-trivial monodromies
by elements of $SL(2,\IZ)$ around singular fibers. Under an $SL(2,\IZ)$ 
transformation
$$ \tau \r { a\tau + b \over c \tau +d}, $$
$\Lambda$ and $\Lambda^*$ transform in the following way, 
\eqn\phase{  \Lambda\, \r \, \Lambda  
\left( {c \bar{\tau} + d \over c \tau +d} \right)^{1/2}
\qquad  \Lambda^* \, \r \,  \Lambda^* 
\left( {c \bar{\tau} + d \over c \tau +d} \right)^{-1/2}. }
Our $H$ and $H'$ field strengths can therefore 
be rotated non-trivially by $SL(2,\IZ)$ 
as we move along the base space $\B$. 

\newsec{Constructing Vacua With $G$-Flux}
\subsec{$K3\times K3$}

The first example that we will consider is M theory on $K3_1\times K3_2$. This
compactification space gives N=4 supersymmetry in three dimensions. 
The anomaly $\chi/24 = 24$
which must be cancelled by a combination of branes and $G$-flux. We therefore
require that,
$$ {1\over 2} \int{ {G\over 2\pi}\wedge {G\over 2\pi}} \leq 24. $$  
Let $J_1$ and $J_2$ denote the \kh\ forms for $K3_1$ and $K3_2$, respectively.
We will initially choose $G$-flux of the form,
\eqn\kthree{{G\over 2\pi} = \omega_1 \wedge \omega_2,}
where $ \omega_i \in H^{1,1}(K3_i, \IZ)$. Each $\omega_i$ must also be primitive
with respect to $J_i$. By definition, each $ \omega_i$ is an element of the 
Picard group $Pic(K3_i)$. 

For a generic $K3$, the Picard group will be empty. To 
find a solution, let us construct a $K3$ surface in the following way: take a
hypersurface in $ \IP^1\times \IP^1 \times \IP^1$ of degree $(2,2,2)$. There are
three natural elements of $Pic(K3)$ which we will denote $C_1, C_2, C_3$. These
classes are determined by the three hyperplanes,
\eqn\hyper{ \eqalign{ \{ p\}\times \IP^1 \times 
\IP^1, \cr 
\IP^1 \times \{ p\} \times \IP^1, \cr
\IP^1 \times \IP^1 \times \{ p\},}}
for $\{ p \}$ a point. With a standard abuse of notation, we will use $C_i$
to denote both the cohomology class and the cycle dual to the Poincar\'e dual of 
$C_i$.  The intersection matrix for the $C_i$ is easily computed. Any two distinct
$C_i$ intersect on a $\IP^1$. A quadratic in $\IP^1$ gives two points. The 
self-intersection number of any $C_i$ vanishes, so we obtain the following 
intersection matrix:
\eqn\intersection{ \pmatrix{0&2&2\cr 2&0&2\cr 2&2&0}.}
We can take $J = C_1+C_2+C_3$ as the \kh\ form for this surface. As our basic 
primitive class, let us take $\a = C_1 - C_2.$ The self-intersection number
of $\a$ is $-4$. 

We can take both $K3_1$ and $K3_2$ to be surfaces of the kind described above.
Each space is endowed with a primitive form denoted $\a_1$ and $\a_2.$ To cancel
the anomaly completely, we can place $16$ membranes on $K3_1\times K3_2$ and turn on,
\eqn\solution{ {G\over 2\pi} = \a_1 \wedge \a_2. }
We can also cancel the anomaly completely without branes in the following way:
in addition to $\a$, let us consider the primitive class $\b=C_1-C_3$ with
self-intersection $-4$. Then $\a \cdot \b = -2$ and we can turn on the flux,
\eqn\completecancel{  {G\over 2\pi} = (\a_1+\b_1) \wedge \a_2.}
Note that this choice of $G$-flux is a primitive $(2,2)$ class with respect
to each of the $\IP^1\times \IP^1$ choices of complex structure. The full N=4
supersymmetry is therefore preserved. 

As a second more exotic example, let us consider the $K3$ surface obtained 
by quotienting
a square $T^4$ with coordinates $(z^1,z^2)$ by,
\eqn\bigpic{ g_1: \, (z^1,z^2) \r (i z^1, - i z^2).}
Under this $\IZ_4$ quotient action, there are no untwisted $(1,1)$ forms. The 
resulting
$K3$ has Picard number $20$ \rshioda. Linear combinations of the twenty 
twisted sector $(1,1)$
forms are therefore integral classes. This implies that combinations of 
$(2,0)$ and $(0,2)$ forms are also integral classes because $H^2(K3,\IZ)$
has $22$ elements. The intersection matrix for 
 these transcendental integral classes is given 
by \rshioda
$$ \pmatrix{2&0\cr 0&2}.$$
For this orbifold case, we note that the untwisted sector holomorphic 
$(2,0)$ form 
\eqn\hol{\gamma=dz^1 dz^2}
satisfies $\int \gamma \wedge {\bar \gamma}=4$.
Let us take both $K3_1$ and $K3_2$ to be $T^4/\IZ_4$
orbifolds. Then the $(2,2)$ form
\eqn\isintegral{  \left\{\gamma_1 \wedge {\bar \gamma}_2 +
{\bar\gamma}_1 \wedge {\gamma}_2 \right\}, }
defined on $K3_1\times K3_2$ is primitive and integral. 

We also require a class $\lambda$, which we take to be a primitive 
$(1,1)$ class with self-intersection $-4$. For example, the class obtained by 
taking the difference of the cycles coming from the resolution of the 
$$ (z^1=1/2,\, z^2=1/2), \quad (z^1=i/2, \, z^2=i/2)$$ 
fixed points in the $(g_1)^2$ twisted sector. This class 
has zero intersection with every other $(1,1)$ form. There are a number of 
other choices for $\lambda$. We can then cancel the anomaly completely
with the following $G$-flux:
\eqn\completeG{ {G\over 2\pi} = \lambda_1\wedge \lambda_2 + 
\gamma_1 \wedge {\bar \gamma}_2 + {\bar\gamma}_1 \wedge {\gamma}_2 .}
Note that this choice of $G$-flux does not preserve the full N=4 supersymmetry.
Varying the complex structure of $K3_i$ rotates $\gamma_i, {\bar\gamma}_i$
and $J_i$ into each other. The resulting $G$ is no longer supersymmetric. Therefore,
only N=2 supersymmetry survives. 
We will contruct another example of a $K3\times K3$ compactification with 
$G$-flux in the following section.    

Before leaving this case, let us see how special choices of 
$G$-flux appear in the $E_8\times E_8$ heterotic dual.
For illustration, let us take the form \kthree\ for our $G$-flux.  
M theory on $K3_1\times K3_2$ has a dual realization in terms of the 
heterotic string on $T^3 \times K3_2$. Away from points of enhanced symmetry,
the heterotic string on $T^3$ has $22$ abelian gauge-fields. As mentioned before,
the gauge-fields
arise in M theory by reducing the $C$-field on the $22$ elements of $H^2(K3, \IZ)$. 
We can then express $G$ reduced on $K3_1$ in terms of the field strength $F$ for
an abelian gauge-field, 
$$ {G\over 2\pi} = F\wedge \omega_1.$$
We then take $F = \omega_2$, which corresponds on the heterotic side to taking 
an abelian connection with some instanton number 
on $K3_2$. Any membranes used to cancel the anomaly correspond to heterotic 
five-branes wrapping
$T^3$. For the $16$ gauge-fields coming from the Cartan of $E_8\times E_8$, 
supersymmetry requires that,
$$ g^{a\bbar} F_{a \bbar} = 0, $$
and that $F$ be a $(1,1)$ form. Clearly these constraints are satisfied by any 
$G$-flux of the form \kthree. 

\subsec{Some orbifold examples with constant $G$-flux}

The next class of examples that we will construct have orbifold singularities.
These examples  all have {\it constant} $G$-fluxes. In cases with F theory
lifts, the corresponding $H$ and $H'$-fluxes will be also be constant. Let
$(z^1,z^2,z^3,z^4)$ coordinatize $T^8$. Since we will consider only $\IZ_2$
quotients, we restrict $T^8$ to $T^2\times T^2 \times T^2\times T^2$ with
each $T^2$ rectangular. We choose each $T^2$ to have periods, 
$$ \int_{\gamma_x^j} dz^i = \delta^i_j
\qquad \int_{\gamma_y^j} dz^i=i\, \delta^i_j, $$
where $\gamma_x^i$ and $\gamma_y^i$ are the $x$ and $y$ one-cycles. 

Since our spaces will be $\IZ_2$ quotients of $T^8$, the metric is flat away
from the singularities and the \kh\ form is, 
\eqn\newkh{J = \sum_i dz^i\wedge d\bar{z}^i.}   
Let us take $G$ to have the form,
\eqn\Gchoice{ \eqalign{ {G\over 2\pi} = & A \, d\zbar^1 dz^2 d\zbar^3 dz^4 + A^* \,  
dz^1 d\zbar^2 dz^3 d\zbar^4 + B \, d\zbar^1 dz^2 dz^3 d\zbar^4  + \cr
& B^* \, dz^1 d\zbar^2 d\zbar^3 dz^4+ C \, d\zbar^1 d\zbar^2 dz^3 dz^4  
+ C^* \, dz^1 dz^2 d\zbar^3 d\zbar^4 .} }
This choice of $(2,2)$ form certainly satisfies $J\wedge G=0$. By construction,
$G$ is real. 
We also require that $ G/2\pi$ be (half)-integer quantized. Requiring that $G/2\pi$
be integral over all four-cycles of $T^8$ gives the conditions:
\eqn\extracond{  2 \left\{ {\rm Re} A \pm {\rm Re} B \pm {\rm Re} C \right\} \in \IZ, 
\qquad  2 \left\{ {\rm Im} A \pm {\rm Im} B \pm {\rm Im} C \right\} \in \IZ.}  
The anomaly condition becomes,
\eqn\flatanom{ 16 \left\{ |A|^2 + |B|^2 + |C|^2 \right\} + n = {\chi \over 24}, }
where $n$ is the number of branes.

Since we will consider orbifolds of $T^8$, we also need to ensure the following two 
conditions: the first is that $G/2\pi$ satisfy the quantization condition \extracond\
modified to take the orbifold action into account.\foot{We wish to thank J. Polchinski
for correcting our original quantization condition.}
The second condition is that $G/2\pi$
has (half)-integer intersections with all 4-cycles coming from the 
twisted sectors.  We will need to check both these conditions on a case 
by case basis,
but one possibility can be removed immediately. Certain twisted sectors can give
rise to operators $\cal O$ which correspond to $(2,2)$ forms. 
However the two-point function of $\cal O$ with $G$ satisfies,
\eqn\bigtwist{ < G \, {\cal O}>=0, }
because ${\cal O}$ is charged under the orbifold gauge group. The remaining 
possibility is three-point functions of the form,
\eqn\twist{ < G \, {\cal P} \,  {{\cal P}}' >, }
where the two-forms $\cal P$ and ${\cal P}'$ carry opposite discrete charge.

As a first example of this kind, let us revisit $K3_1\times K3_2$ where we
realize $K3_i$ by $T^4/\IZ_2$. We therefore quotient $T^8$ by $\IZ_2\times \IZ_2$
generated by,
$$ \eqalign{ & g_1: \, (z^1,z^2) \r (-z^1, -z^2) \cr
             & g_2: \, (z^3,z^4) \r (-z^3, -z^4). }$$
The form \Gchoice\ is invariant under this action. The first issue is how the 
quantization condition is modified. The volume of the fundamental domain for the orbifold
has volume reduced by $1/4$. For example, we can parametrize the fundamental domain by 
restricting the range of $x_1$ and $x_3$ to be $0$ to $1/2$ rather than $0$ to $1$. 
There are now $4$-cycles
with $1/4$ the volume. These give a modified quantization condition, 
\eqn\newmodquant{{1\over 2} 
\left\{ {\rm Re} A \pm {\rm Re} B \pm {\rm Re} C \right\} \in \IZ, 
\qquad  {1\over 2} \left\{ {\rm Im} A \pm {\rm Im} B \pm {\rm Im} C \right\} \in \IZ,}
and anomaly constraint:
\eqn\modflatanom{ 4 \left\{ |A|^2 + |B|^2 + |C|^2 \right\} + n = {\chi \over 24}. }
For appropriate choices
of $A, B$ and $C$, we obtain supersymmetric compactifications. For example,
the choice
$$ A = 2 \qquad B = 1 \qquad C=1,$$
cancels the anomaly without any branes. 
To check that $G/2\pi$ is an integer form, we also need to compute \twist.
First we note that $G\wedge \omega_1  = G\wedge \omega_2=0$ for $\omega_i$
the volume form of $K3_i$.  This guarantees that if 
$ {\cal P}$ and ${\cal P}'$ are two-forms from the same $K3$, \twist\
vanishes. The remaining possibility is when 
$ {\cal P}$ and  ${\cal P}'$ are charged under different $\IZ_2$ actions in
which case \twist\ vanishes by charge conservation. 

A more interesting example is the $\IZ_2$ quotient by the action,
$$ g_1: \, (z^1,z^2, z^3, z^4) \r (-z^1, -z^2, -z^3, -z^4). $$
This space has singularities which cannot be resolved. However, it is 
perfectly fine as an M theory or type IIA compactification. String orbifold 
techniques
give $\chi/24 = 16$. The Hodge numbers are: 
$$H^{2,0}=6, \quad
H^{1,1}=16, \quad H^{2,1}=0, \quad H^{3,1}=16, \quad H^{2,2} = 292.$$
As usual, $H^{4,0}=1$. There are now $4$-cycles with volume reduced by $1/2$. 
These give the following quantization condition:
\eqn\newtmodquant{ 
\left\{ {\rm Re} A \pm {\rm Re} B \pm {\rm Re} C \right\} \in \IZ, 
\qquad   \left\{ {\rm Im} A \pm {\rm Im} B \pm {\rm Im} C \right\} \in \IZ,}
and anomaly constraint:
\eqn\modtflatanom{ 8 \left\{ |A|^2 + |B|^2 + |C|^2 \right\} + n = {\chi \over 24}. } 
Tuning $A, B$ and $C$ appropriately gives solutions
that cancel the tadpole either partially or completely. 
For example, the choices $ A=1+i$ or $A=B=1$ both completely cancel 
the tadpole with just $G$-flux. 
Checking that $G/2\pi$ is integer is easy in this case because there are 
no twisted sector two-forms $\cal P$. The only operators are four-forms $\cal O$
whose intersection with $G/2\pi$ vanishes. 

Our next example is the symmetric product of $K3$. It is of a
quite different flavor because we will {\it not} be able to find a
supersymmetric solution with flux of the form \Gchoice. We define the quotient
by the action,
$$ \eqalign{ & g_1: \, (z^1,z^2) \r (-z^1, -z^2) \cr
             & g_2: \, (z^3,z^4) \r (-z^3, -z^4) \cr
             & g_3: \, (z^1, z^2, z^3, z^4) \r (z^3, z^4, z^1, z^2). }$$
This compactification space has N=3 supersymmetry in three dimensions because $S^2(K3)$
is a hyper\kh\ space. In this case, $\chi/24 = 27/2$. The Hodge numbers are: 
$$H^{2,0}=1, \quad  
H^{1,1}=21, \quad H^{2,1}=0, \quad H^{3,1}=21, \quad H^{2,2} = 232.$$
To obtain a 
consistent compactification, we therefore need to turn on $G$-flux. Note that
invariance under the $g_3$ action requires that $B$ and $C$ be real.
In this case, $G/2\pi$
can be half-integer quantized \rfluxquant.
At first sight it seems that this
additional freedom is not enough to find a solution to \flatanom. Let us begin by
examining the integrality condition on $G/2\pi.$ For the moment, let us ignore the
effect of quotienting by $g_3$ and start by considering just $K3 \times K3$.
The integrality condition becomes,  
\eqn\newcond{   \left\{ {\rm Re} A \pm  B \pm C \right\} \in \IZ, 
\qquad   \left\{ {\rm Im} A \right\} \in \IZ. }
Including the effect of symmetrizing can only make this condition more stringent. 
The anomaly constraint (including the effect of symmetrizing) becomes, 
\eqn\symmanom{ 2 \left\{ |A|^2 + B^2 + C^2 \right\} + n = {27 \over 2}. } 
 

Let us derive the condition that the
intersection of $G/2\pi$ with all twisted sector states is half-integral. 
The intersections with the twisted sector states corresponding to the generators
$g_1$ and $g_2$ vanish by exactly the arguments
given for the $T^8/(\IZ_2)^2$ example above. We therefore only need
to check the condition for twisted sector states generated by $g_3$. 
For this purpose, it is easier to
consider $S^2(T^4)$ rather than $S^2(K3)$. This simplification is possible
because the unique twisted sector $(1,1)$ form  on
$S^2(K3)$ descends from the unique twisted sector $(1,1)$ form on $S^2(T^4)$.

Denote $S^2(T^4)$ by $X$ and let ${\widetilde X}$ be its resolution 
obtained by blowing up over
the fixed locus which is the diagonal four-torus $T^4_D$.\foot{We wish to 
thank D.-E. Diaconescu for explaining the following argument.} Let $Y=T^4\times T^4$
and ${\widetilde Y}$ be the space obtained by blowing up $Y$ over the diagonal
$T^4_D$. There are projections $q$ and $p$ from ${\widetilde Y}$ to $Y$ 
and ${\widetilde X}$
to $X$ respectively. The involution $i: Y \rightarrow X$ lifts to an 
involution $s: {\widetilde Y} \rightarrow {\widetilde X}$, which is branched over 
the exceptional divisor.
We can summarize this information in the following commutative diagram:
\eqn\commutative{\matrix{ {\widetilde Y} & \matrix{s \cr \longrightarrow \cr
\phantom{X}}& {\widetilde X}\cr
\downarrow q &{}&\downarrow p\cr
 Y &\matrix{i\cr \longrightarrow\cr
\phantom{X}}& X.}}
Now consider a form $\omega \in H^{2,2}(Y) \, \cap \, H^4(Y, \IZ)$. For
example, the form defined in \Gchoice. This form can be pulled back
to $ \widetilde{Y}$ giving $q^*\omega \in H^{2,2}({\widetilde Y})$ and then
pushed forward to a form:
$${\widetilde \omega} = s_*(q^*\omega) \in H^{2,2}({\widetilde X}).$$
Note that ${\widetilde Y}$ is a double cover of ${\widetilde X}$.
Therefore $s^*(s_* v )= 2v $ for any form $v$ on ${\widetilde Y}$ and we have
$s^*{\widetilde \omega} = 2q^*\omega$. 
Consider the integral twisted sector
$(1,1)$ form $\widetilde{t}\in H^{1,1} ({\widetilde X})\, \cap \,
H^4({\widetilde X}, \IZ)$ which is Poincar\'e dual to the exceptional divisor
in $ \widetilde X.$ Then
$$s^*{\widetilde t} = 2t,$$ 
where  $t\in H^{1,1} ({\widetilde Y})\, \cap \, H^4({\widetilde Y}, \IZ)$ is
Poincar\'e dual to the exceptional divisor in $\widetilde Y$. This is true
again because $\widetilde Y$ is a double cover of $\widetilde X$ and $s$ is
branched over the exceptional divisor. 
The relevant three point function we wish to compute is
$$<{\widetilde t}^2 \cdot {\widetilde \omega}>_{\widetilde X} \, = \,
{1\over 2} <s^*({\widetilde t}^2 \cdot {\widetilde \omega})>_{\widetilde Y}
\,=\,
2<t^2\cdot s^*{\widetilde \omega}>_{\widetilde Y} \,=\,
4<t^2\cdot q^*{\omega}>_{\widetilde Y}.$$
Since the normal bundle of the diagonal $T^4_D$ is trivial, we see that
the exceptional divisor $D$ corresponding to $t$ is just $\IP^1\times T^4_D$.
The intersection is then
$$-8 \, <T^4_D\cdot q^*\omega>_{\widetilde Y} \, = -8\, 
<T^4_D\cdot \omega>_Y.$$
But this can now be readily computed: first it is easy to check that 
the contributions from the $A$ and $A^*$ terms
vanish because the forms themselves vanish on $T^4_D$. The other terms yield
the following integrality condition for half-integral $G$-flux,
$$ 128(B+C) \in \IZ.$$
While this was derived for $S^2(T^4)$, the same condition can be seen to
hold for the symmetric product of $K3$ in the orbifold limit.\foot{The
symmetric product is 
a highly singular space. However, we can smooth the space
by blowing up the symmetric product $X=S^2(K3)$ over the fixed locus.  This amounts 
to replacing $X$ by the Hilbert scheme 
of points $ K3^{[2]}$. See, for example, \rnaka. In a diagram analogous to 
\commutative, $ \widetilde{X} =  K3^{[2]}$ and $p: K3^{[2]} \r X$. 
 If we want  $ p^*( G/\pi)$ to be a primitive 
element of $H^{2,2}(K3^{[2]}, \IZ)$, we need to 
impose at least one additional condition
on \Gchoice. The \kh\ class of $ K3^{[2]}$ has a term proportional 
to the class of the exceptional divisor. To ensure primitivity, we can require
that $G/\pi$ vanish on the fixed locus, which implies that $B=C$. 
This is a natural way to get a smooth hyper\kh\ compactification. 
We wish to thank L. G\"ottsche for pointing out
this generalization.}

What remains is for us to find a solution to these integrality conditions, and here
we meet a problem. Even the weaker condition \newcond\ requires that
one of ${\rm Re A},B$ or $C$ be integral. For example, we can choose
$$ A = a+i\widetilde{a}, \qquad B=b/2, \qquad C=c/2, $$
where $a, {\widetilde a}, b, c$ are integers, and where 
$b,c$ are both even or both odd.
However, the anomaly cancellation condition imposes the constraint:
$$ a^2 + \widetilde{a}^2 + {b^2 + c^2 \over 4} + {n\over 2} = 6+{3\over 4}.$$ 
This condition cannot be satisfied! There is therefore no choice of flux
of the form \Gchoice\ which preserves supersymmetry in this example. 


Lastly, we consider an example which gives N=2 supersymmetry in three dimensions.
We quotient by, 
$$ \eqalign{ & g_1: \, (z^1,z^2) \r (-z^1, -z^2) \cr
             & g_2: \, (z^1,z^3) \r (-z^1, -z^3) \cr
             & g_3: \, (z^2,z^4) \r (-z^2, -z^4).} $$
The quotient group essentially inverts all possible pairs of tori. For this case,
$\chi/24 = 28$. The Hodge numbers are: 
$$H^{2,0}=0, \quad H^{1,1}=100, \quad H^{2,1}=0, \quad H^{3,1}=4, 
\quad H^{2,2} = 460.$$ 
The integrality condition becomes, 
\eqn\lastcase{   {1\over 4} \left\{ {\rm Re} A \pm {\rm Re}  B \pm {\rm Re} 
C \right\} \in \IZ, 
\qquad  {1\over 4}  \left\{ {\rm Im} A  \pm {\rm Im} B \pm {\rm Im} C\right\} \in \IZ, }
with anomaly constraint,
\eqn\symmanom{  2 \left\{ |A|^2 + |B|^2 + |C|^2 \right\} + n = 28. } 
As a sample choice of $G$-flux, we can take
$$ A= 2, \qquad B= 2, \qquad C=0, $$
which cancels the anomaly completely with $n=12$ branes. To check that $G/2\pi$
is integer, we need to check that it has integer periods over all integer
homology cycles. For the untwisted sector cycles, this reduces to
verifying \lastcase, which is obvious. For the twisted sector cycles, it is
easy to repeat the arguments presented for the previous $T^8/(\IZ_2)^2$ example
to show that all three point functions of type \twist\ vanish. 

\subsec{An orientifold example}

In a similar way, we can construct four-dimensional orientifold examples with 
constant fluxes. 
We start with an orbifold of the form $T^6/\Gamma$. 
Again for simplicity let us take $T^6 = T^2\times T^2\times T^2$ with each
factor rectangular and coordinates $(z^1,z^2,z^3)$. We can view an orientifold
of $T^6/\Gamma$ as a special point in the moduli space of F theory compactified
on the elliptically-fibered four-fold \rsenftheory, 
$$ {\M} =  {T^6/\Gamma \times T^2 \over {\IZ_2} }, $$
where the $\IZ_2$ action inverts both 
$z^3$ and the coordinate, $z^4$, of the fiber $T^2$. This 
F theory compactification reduces 
to the orientifold of type IIB on $T^6/\Gamma$ by the action  
$ \Omega (-1)^{F_L} \IZ_2$ where the $\IZ_2$ inverts $z^3$. This action
produces various $O7$-planes at complex codimension one fixed sets on $T^6$. 
By adding $D7$-planes, we can cancel the $O7$-plane charge. Fixed points
under group elements in the product of $ \Omega (-1)^{F_L} \IZ_2$ and
$\Gamma$ generate $O3$-planes. The total $D3$-brane tadpole is $\chi({\M})/24$.

As a specific example, we can take the following generators for $\Gamma$,
$$ \eqalign{ & g_1: \, (z^1,z^2) \r (-z^1, -z^2) \cr
             & g_2: \, (z^1,z^3) \r (-z^1, -z^3).}$$
The resulting Calabi-Yau $T^6/\Gamma$ has Hodge numbers $h^{1,1}=51, 
\, h^{2,1}=3$. The associated four-fold $\M$ is the final example of
section {\it 3.2}. This orientifold and its relation to F theory 
has been studied in \refs{\rgopak, \rzurab}. 
To cancel the anomaly with $12$ $D3$-branes, we can
turn on the fluxes
\eqn\oriflux{ \eqalign{ H & =   A \, d\zbar^1 dz^2 d\zbar^3  + A^* \,  
dz^1 d\zbar^2 dz^3  + B \, d\zbar^1 dz^2 dz^3  + 
B^* \, dz^1 d\zbar^2 d\zbar^3,\cr
             H' & = A i \, d\zbar^1 dz^2 d\zbar^3 - A^* i\,  
dz^1 d\zbar^2 dz^3  - B i\, d\zbar^1 dz^2 dz^3   + 
B^* i \, dz^1 d\zbar^2 d\zbar^3,} }
with the choice:
$$ A = 2,  \qquad B=2.$$
It is interesting to note that the same four-fold can give rise to many
different orientifolds depending on the choice of $C$ and $G$-flux. This
point will be explored more fully elsewhere.

\newsec{A Heterotic Compactification With Torsion}

In this section, we will construct an example of a four-dimensional
$SO(32)$ heterotic string compactification with torsion. This particular
example has either N=1 or N=2 spacetime supersymmetry, depending on the choice
of flux. We begin with type IIB  
compactified on an orientifold of $K3\times T^2$. After 
a series of $T$ and $S$ dualities, we will arrive at our heterotic model. 

The initial IIB supergravity metric is conformal to the metric on $K3\times T^2$. 
The solution still possesses two isometries along the $T^2$. Two T-dualities 
along the two circles of $T^2$ sends
$\Omega (-1)^{F_L} \IZ_2$ to $\Omega\,$ \rsenorien. In other words, it takes our 
F theory compactification with fluxes to a type I theory. 
In the subsequent discussion, we will specify the resulting type I background.
If we choose to use only $D3$-branes and no background flux to cancel the anomaly,
the resulting theory is type I on $K3\times T^2$ with 24 
$D5$-branes wrapping the $T^2$. With background fluxes, the result is quite
different. We will find type I compactified on a space $\B'$ with the 
following properties:

1. It is a complex manifold which is not \kh, or even conformally Calabi-Yau.

2. It has vanishing first Chern class.

3. It has a non-zero $H'$-flux.

\noindent After a further $S$-duality, we end up with the $SO(32)$ 
heterotic string compactified on $\B'$ with a non-zero $H$-flux. 

\subsec{Mapping the parameters and couplings}

Let us consider the orientifold of type IIB on $K3\times T^2$ 
with a square $T^2$ by the
action  $\Omega (-1)^{F_L} \IZ_2$. The $\IZ_2$ action sends $z^1\r -z^1$
where $z^1$ is the coordinate of the $T^2$. This compactification is a special 
point in the moduli space of F theory on $K3 \times K3$.
Let the $T^2$ have sides of length 
$R_1$ and $R_2$, and volume $\widetilde{V} = R_1 R_2$. We will take the $K3$
to have volume $V$. At the orientifold point, the ten-dimensional 
string coupling is 
a free parameter which we will take to be $g_B$. 

Two T-duality transformations along $T^2$ invert the radii in the
usual way,
$$ R_i \, \r \, \alpha'/R_i,$$ 
where $i=1,2$. The resulting type I theory has the following couplings
and volumes:
\eqn\coupI{\eqalign{ g_I^{(4)}= g_B/\sqrt{V\widetilde V},\qquad  
g_I^{(10)}= \alpha'g_B/\widetilde{V}, \qquad 
\widetilde{V}_I=\alpha'^2/\widetilde{V}, \qquad &  V_I = V.}}
Here $g^{(4)}$ and $g^{(10)}$ denote the four and ten-dimensional couplings. 
The $O7$-planes and $D7$-branes are mapped to an $O9-D9$ system. If, for simplicity,
we assume a trivial seven-brane gauge-field configuration over $K3_1$ 
then the initial gauge group is  $SO(8)^4$. The gauge-fields of the
resulting $O9-D9$ theory then have non-trivial Wilson lines. 
Under a further $S$-duality transformation, we get the heterotic $SO(32)$ theory. 
The resulting couplings and volumes can again be written in terms of IIB variables,
\eqn\couphet{g^{(4)}_{het}=\sqrt{g_B/(V\alpha')}, \quad 
g^{(10)}_{het}=\widetilde{V}/(g_B\alpha'),\quad 
\widetilde{V}_{het}=\alpha'/g_B, \quad 
V_{het}= V\widetilde{V}^2/(g_B^2\alpha'^2).}

Our initial IIB supergravity description is valid in the limit where,
\eqn\superlimit{ \widetilde{V}/\alpha', \,\, V/(\alpha')^2 >>1.}
If we want a weakly coupled orientifold theory, we can also take $g_B$
to be small but that is not necessary. Under condition \superlimit,
both the heterotic and type I four-dimensional couplings can be made small.
If $g_B$ is small, then an $\alpha'$ expansion of the resulting heterotic string
theory is a good approximation. 

\subsec{The type IIB solution}

To obtain the type IIB supergravity metric,
we begin with M theory on ${\M} = K3_1\times T^4/\IZ_2$. If we choose a smooth $K3_1$
then our resulting type IIB metric will be smooth. We can also consider orbifold
cases where $K3_1=T^4/\Gamma$ where the metric is explicitly known. Otherwise, our 
resulting heterotic solution is given in terms of the metric of $K3_1$. We will use
coordinates $w^a$ for $K3_1$. We again 
take $T^4 = T^2\times T^2$ with coordinates $(z^1,z^2)$ and each factor 
square. Our initial
M theory metric is then of the form \wmetric\ with $g$ the metric on $\M$. We can
choose to completely or partially cancel the anomaly with $G$-flux satisfying,
$$ {1\over 2} \int_{\M} {G\over 2\pi} \wedge {G\over 2\pi} + n = 24. $$  
Using arguments along the lines discussed in section {\it 3}, we can
construct a $G$ with the form
\eqn\Gsol{ {G\over 2\pi} = \a \wedge dz^1 d\zbar^2 + \a^* \wedge d\zbar^1 dz^2 
+ \b \wedge d\zbar^1 d\zbar^2 + \b^*  \wedge dz^1 dz^2,}
where $\a \in H^{1,1}(K3_1)$ and $ \b \in H^{2,0} (K3_1).$ Note that if $\b=0$,
the model has N=2 supersymmetry.

We treat the $z^2$ direction as the elliptic fiber, and lift this M theory vacuum 
to a type IIB orientifold of $K3_1\times T^2$. 
The resulting background 
fluxes are given by,
\eqn\solH{\eqalign{H & = (\a + \b^*) \wedge dz^1 + (\a^*+\b) \wedge d\zbar^1 \cr 
H' &= (\bar{\tau} \a + \tau \b^*) \wedge dz^1 
+ (\tau \a^* +\bar{\tau} \b )\wedge d\zbar^1.}} 
We note that $H$ and $H'$ can be expressed in the form,
\eqn\potential{\eqalign{ H &= d\left\{ \Lambda_{\a+ \b^*} \wedge dz^1
+ \Lambda_{\a^*+\b}  \wedge d\zbar^1 \right\} \cr
 H' &= d\left\{ \Lambda_{ \bar{\tau} \a + \tau \b^*} \wedge dz^1
+ \Lambda_{\tau \a^* +\bar{\tau} \b }  \wedge d\zbar^1 \right\}.}}
The potentials $\Lambda_\gamma$ are not globally defined forms on the space 
$K3_1\times T^2$, but satisfy $ d\Lambda_\gamma = \gamma.$

In string frame, the type IIB supergravity metric has the form:
\eqn\ansatz{\pmatrix{ \Delta'~ \eta_{\mu\nu} &0 \cr
0 & \Delta~ \widetilde{g}.}}
The indices $\mu,\nu=0,\ldots,4$ and $\widetilde{g}$ is the metric of 
$K3_1\times T^2/\IZ_2$.  The warp factors $\Delta$ and $\Delta'$ depend
on the internal coordinates. 
We can determine $\Delta$ and $\Delta'$ in the following way. Let us reduce
from M theory to type IIA along a side of the elliptic fiber. The metric of the
torus is warped,
$$ e^{\phi/2} dz^2 d\zbar^2. $$ 
The resulting type IIA metric in string frame is given by \rWdynm,
\eqn\stringframe{ g_{\rm IIA} = e^{\phi/4} G^{(10)},}
where $G^{(10)}$ is the straight dimensional reduction of the M theory metric.
Using the metric \wmetric, we find that:
\eqn\detmet{ \Delta = ( \Delta')^{-1} =  e^{3\phi/4}.}

Lastly, let us recall that the warp factor is determined by equation \warp. 
There are three source terms on the right hand side of this equation. Both the 
$X_8$ curvature term and the membrane term are suppressed by six powers of 
$M_{pl}$ relative to 
the $G\wedge G$ source term. To leading order in the derivative expansion, we
can neglect the effect of both terms.\foot{ It is 
worth noting that there is an obstruction to solving the warp factor equation \warp.
However, if the anomaly cancellation condition \condition\ is satisfied then the
obstruction vanishes. The $X_8$ and membrane terms are then crucial for 
ensuring the existence of a solution for the warp factor.}
With the form of $G$-flux given in \Gsol, the warp factor will have
no dependence on $(z^1, z^2)$ at the level of the supergravity solution.

We can see this directly in type IIB supergravity. The only non-vanishing
component of $D^+$ is given by,
$$ D^+_{\mu\nu\rho\lambda} = \epsilon_{\mu\nu\rho\lambda} e^{-3\phi/2}. $$
The self-dual field strength $F^+$ given in \Gfive\ obeys \rDJM,
\eqn\bianchi{ \eqalign{ d* F^+  = & H\wedge H' + 
{(4\pi^2 \alpha')^2} \bigg\{ {1\over 64 \pi^2} \sum_{i=1}^4 
\tr (R\wedge R) \, \delta^2(z^1 - z^1_i)
\cr & + \sum_{j=1}^n \delta^2(z^1-z^1_j) \delta^4 (w -w_j) \bigg\} ,}}
where $z_i^1$ are the locations of the $O7$-plane plus four $D7$-branes, and
$(z^1_j, w_j)$ are the locations of the $D3$-branes.
From \bianchi, we obtain an equation for the warp factor:
\eqn\anotherwarp{ \eqalign{ d* d D^+ = &   H\wedge H' +
{(4\pi^2 \alpha')^2} \bigg\{  {1\over 64 \pi^2}\sum_{i=1}^4  \tr (R\wedge R) \,
 \delta^2(z^1 - z^1_i) \cr &
+ \sum_{j=1}^n   \delta^2(z^1-z^1_j) \delta^4 (w -w_j) \bigg\}.}}
Again the $H\wedge H'$ term is constant in $z^1$ while the 
remaining source terms are suppressed by powers of $\alpha'$.

\subsec{Dualizing to a type I solution}

To arrive at a type I solution, we T-dualize along both sides of the $T^2$
with coordinate $z^1 = x^1+iy^1$. Using Buscher's duality \rBUSH\ and \refs{\rKKL, \rBHO,
\rBBO}, we can
express the type I metric $ g^I $ and RR two-form ${B'}^I$ in terms of our
initial type IIB quantities. 
Since $H$ and $H'$ can be expressed in the form \potential, 
we only have $B_{xa}, B_{xa}'$ and $B_{ya}, B_{ya}'$
components.
The type I metric is then given by,\foot{We set $4\pi^2 \alpha'=1$ for the remainder of
the paper.}
\eqn\typeimetric{ \eqalign{g_{ab}^I  = \Delta \widetilde{g}_{ab} + 
{1\over \Delta \widetilde{g}_{xx}} B_{xa} B_{xb} +
{1\over \Delta \widetilde{g}_{yy}} B_{ya} B_{yb}, \cr
g_{xa}^I  = {1\over \Delta \widetilde{g}_{xx}} B_{xa}, \qquad
g_{ya}^I  = {1\over \Delta \widetilde{g}_{yy}} B_{ya}, \cr
 g_{xx}^I = {1\over \Delta \widetilde{g}_{xx}}, \quad
g_{yy}^I  = {1\over \Delta \widetilde{g}_{yy}}, \quad g_{xy}^I  = 0,\cr }}
where the $a, b$ directions are along $K3_1$. 
The type I dilaton is inversely proportional to the warp factor:
\eqn\dilatonI{ e^{\phi^I} = {g_B\over \Delta \sqrt{\,\widetilde{g}_{xx} 
\widetilde{g}_{yy}}}. }
The ${B'}^I$-field is given by,
\eqn\Bone{ \eqalign{ 
{B'}^I_{ab}= {3\over 2}\left\{ B_{x [a} B'_{by]} - B_{x [a}' B_{by]}
\right\} + 2 B'_{x[a} B_{b]y}, \cr
{B'}^I_{xa}=  B'_{a y}, \qquad {B'}^I_{ya}= -B'_{a x}, \qquad
{B'}^I_{xy}=0. \cr}}
Note that we have used the fact that $D^+$ has no internal components in the
above expressions. If we set either $\a$ or $\b$ to zero in \solH, then
one can check using the local expressions \potential\ 
for $B$ and $B'$ that ${B'}^I_{ab}=0$.

Using coordinates $w^a$ for $K3_1$, we can rewrite the type I metric in a way 
that makes the structure a little clearer:
\eqn\rewrite{ ds^2 =  \Delta \widetilde{g}_{ab} dw^a dw^b
+  {1\over \Delta \widetilde{g}_{xx}} \left( dx + B_{xa} dw^a \right)^2 +
 {1\over \Delta \widetilde{g}_{yy}} \left( dy + B_{ya} dw^a \right)^2.}
The $T^2$ parametrized by $x$ and $y$ is now non-trivially twisted
over the base $K3_1$. By viewing $B_{xa} dw^a$ and $B_{ya} dw^a$ as 
Kaluza-Klein gauge-fields, we see that the twisting is encoded in the 
characteristic classes of these gauge-fields on $K3_1$.

Since T-duality is a perturbative symmetry, we 
have arrived at a 
consistent type I background \typeimetric, \dilatonI\ and \Bone. We can S-dualize to
a perturbative heterotic compactification with dilaton,
\eqn\hetdilaton{ e^{\phi_{het}} = {1\over g_B} \, \Delta \sqrt{\,\widetilde{g}_{xx} 
\widetilde{g}_{yy}},}
and string frame metric:
\eqn\hetfields{ {1\over g_B} \sqrt{\,\widetilde{g}_{xx} 
\widetilde{g}_{yy}} \pmatrix{
\eta_{\mu \nu} & 0 \cr 0 & \Delta \, {g}^I }. }
Under S-duality, $ {B'}^I \, \r \, B^{het}$. 

Torsional string
compactifications satisfy a number of stringent constraints. See \refs{\rrocek,
\rkirit, \rdewit, \rstrom, \rhull} for a discussion of various aspects of 
torsional compactifications. As a final comment, we note that
since our torsional vacuum is T-dual 
to an anomaly 
free orientifold, we can infer a great deal about the resulting
metric and $B$-field. Our space should have  trivial
first Chern class.  Since no  $SO(32)$ gauge-fields are excited,  
$p_1(\B')$ must also be trivial in cohomology to satisfy the usual
anomaly cancellation condition,
$$ dH^{het} =  \tr(R\wedge R)-{1\over 30} \Tr (F\wedge F).$$
This is not as implausible as it might first seem. In the $E_8\times E_8$
heterotic dual to the original type IIB orientifold, $p_1(K3)$ is
cancelled by embedding instantons in some abelian gauge-fields. It seems
reasonable that the instantons in the Kaluza-Klein gauge-fields  
$B_{xa} dw^a$ and $B_{ya} dw^a$ could analogously cancel $p_1(K3)$ in this case. 
Finally, it should be possible to show that the metric \hetfields\ and 
$H^{het}$ are derivable from a real prepotential, 
analogous to the \kh\ potential \rrocek.

\bigbreak\bigskip\bigskip\centerline{{\bf Acknowledgements}}\nobreak

It is our pleasure to thank S. Mukhi for early participation, and
D. R. Morrison for late participation in this project. We
are particularly grateful to D.-E. Diaconescu and  D. R. Morrison 
for numerous suggestions and comments. We have also been aided by conversations with
L. G\"ottsche, Z. Kakushadze, 
A. Klemm, M.  Ro\v cek,  A. Sen and E. Witten. 
The work of K.D. is supported in part by DOE Grant No. DE-FG02-90-ER40542. 
The work of R.G. is supported in part
by NSF grant No. DMS-9627351 and by NSF Grant No. PHY94-07194, while that of 
S.S. is supported  by the William Keck Foundation and by 
NSF grant No. PHY--9513835.

\vfill\eject

\listrefs
\bye